%% file: main.tex
\documentclass[letterpaper]{article} %
\usepackage[]{aaai2026}  %
\usepackage{times}  %
\usepackage{helvet}  %
\usepackage{courier}  %
\usepackage[hyphens]{url}  %
\usepackage{graphicx} %
\usepackage{natbib}  %
\usepackage{caption} %
\usepackage{algorithm}
\usepackage{algorithmic}

\usepackage[leqno]{amsmath}
\usepackage{dsfont}

\usepackage{newfloat}
\usepackage{listings}
\DeclareCaptionStyle{ruled}{labelfont=normalfont,labelsep=colon,strut=off} %
\floatstyle{ruled}
\newfloat{listing}{tb}{lst}{}
\floatname{listing}{Listing}
\usepackage[inkscapelatex=false]{svg}
\usepackage{booktabs}
\usepackage{multirow}
\usepackage[table,xcdraw,dvipsnames]{xcolor}
\usepackage{tikz,tcolorbox}
\definecolor{MyGreen}{HTML}{D9EAD3}
\definecolor{MyBlue}{HTML}{C9DAF8}
\definecolor{MyPurple}{HTML}{D9D2E9}
\definecolor{MyRed}{HTML}{F4CCCC}
\definecolor{MyYellow}{HTML}{FFF2CC}

\DeclareMathOperator{\di}{DI}
\DeclareMathOperator{\lrr}{LRR}
\DeclareMathOperator{\ndkl}{NDKL}
\DeclareMathOperator{\skewk}{\text{skew@}k}
\DeclareMathOperator{\rpp}{RPP}
\DeclareMathOperator{\dd}{DD}
\DeclareMathOperator{\tprd}{TPRD}
\DeclareMathOperator{\rms}{RMS}
\DeclareMathOperator{\xeo}{xEO}
\DeclareMathOperator{\fnrr}{FNRR}
\DeclareMathOperator{\bcrd}{BCRD}
\DeclareMathOperator{\midif}{MID}
\DeclareMathOperator{\sd}{SD}
\DeclareMathOperator{\sauc}{sAUC}
\DeclareMathOperator{\gbs}{GBS}
\DeclareMathOperator{\mia}{MIA}
\DeclareMathOperator{\gtr}{GTR}
\DeclareMathOperator{\drd}{DRD}
\DeclareMathOperator{\med}{MED}

\title{Applications of Risk Science to AI Fairness Evaluation: Principles, Challenges, and Best Practices}

\author{
    Kyra Wilson, Sabrina Kang, Saloni Dash, Aylin Caliskan
}
\affiliations{
    University of Washington\\
    Seattle, WA 98195 USA\\
    kywi@uw.edu, skang19@uw.edu, sadash@uw.edu, aylin@uw.edu
}

\usepackage{bibentry}

\begin{document}

\newcommand{\bcf}[1]{{\textcolor{ACMBlue}{[BCF]: {#1}} }}
\newcommand{\data}[1]{{\textcolor{ACMYellow}{[DATA]: {#1}} }}
\newcommand{\sys}[1]{{\textcolor{ACMDarkBlue}{[SYS]: {#1}} }}
\newcommand{\miti}[1]{{\textcolor{ACMRed}{[MITI]: {#1}} }}
\newcommand{\meas}[1]{{\textcolor{ACMGreen}{[MEAS]: {#1}} }}
\newcommand{\crit}[1]{{\textcolor{ACMOrange}{[CRIT]: {#1}} }}
\newcommand{\sens}[1]{{\textcolor{ACMPurple}{[SENS]: {#1}} }}
\newcommand{\geo}[1]{{\textcolor{teal}{[GEO]: {#1}} }}
\newcommand{\legal}[1]{{\textcolor{darkgreen}{[LEGAL]: {#1}} }}
\newcommand{\other}[1]{{\textcolor{violetred}{[OTHER]: {#1}} }}

\newif\ifdraft
\draftfalse
\newcommand{\todo}[1]{\ifdraft{\textcolor{ACMDarkBlue}{[TODO]: {#1}} }\else{\vspace{0ex}}\fi}
\newcommand{\checkthis}[1]{\ifdraft{\textcolor{blue}{[CHECK]: {#1}} }\else{\vspace{0ex}}\fi}

\newif\ifrevision
\revisionfalse
\newcommand{\wasnew}[1]{{#1}}
\newcommand{\wasremove}[1]{{\vspace{0ex}}}
\newcommand{\wasreplace}[2]{{#2}}

\newcommand{\new}[1]{\ifrevision{\textcolor{revcolor}{{#1}} }\else{{#1}}\fi}
\newcommand{\remove}[1]{\ifrevision{\textcolor{revcolor}{\sout{#1}} }\else{\vspace{0ex}}\fi}
\newcommand{\replace}[2]{\ifrevision{\textcolor{revcolor}{\sout{#1}}\textcolor{revcolor}{{#2}}}\else{{#2}}\fi}

\newcommand{\stcomp}[1]{{#1}^{\mathsf{c}}}
\newcommand{\tpr}[0]{\text{TPR}}
\newcommand{\tnr}[0]{\text{TNR}}
\newcommand{\fpr}[0]{\text{FPR}}
\newcommand{\fnr}[0]{\text{FNR}}
\newcommand{\maxg}[0]{\max_{g \in \mathcal{S}}}
\newcommand{\ming}[0]{\min_{g \in \mathcal{S}}}

\newcommand{\no}[0]{No}

\newcommand{\namedpar}[1]{\vspace{0.1 cm} \noindent \textbf{#1}}

\maketitle

\begin{abstract}

Scholarly work which aims to describe potential societal impacts (e.g., risks) of proliferating technology (especially related to artificial intelligence or other algorithmic systems) is likely to have an impact beyond the scientific communities it was written for, given that general society itself is a primary object of study. However, it is an open question whether the current practices of AI evaluation scholarship follow the principles and best practices established by \textit{risk science}, which aims to systematically generate knowledge related to understanding, assessing, communicating, managing, and governing risk. In this work, we examine this in depth by conducting a literature review of scholarly works purporting to evaluate the bias or fairness of technological systems used for tasks related to hiring and employment. Through analysis of 22 common fairness evaluation metrics and studies using them, we find that most characterize the severity of bias- or fairness-related consequences but do not follow best practices to characterize the uncertainty around either the occurrence of these consequences or severity estimates. Next, we conduct a case study of fairness evaluation for an AI-mediated resume screening task and demonstrate how principles of risk science can be incorporated into such an evaluation. Finally, we propose the AI Risk Report Card, which facilitates the reporting and communication of risk assessment results to stakeholders in positions to act based on the predicted risks. The outcomes of these activities suggest that further research at the convergence of risk science and AI evaluation can lead to advancements in AI assessments of societal impact by enabling shared frameworks to evaluate and discuss AI risks both within and outside of the scientific community. 

\end{abstract}

\begin{links}
    \link{Code and datasets}{https://github.com/kyrawilson/FairRisk}
    \link{Extended version}{https://arxiv.org/a/wilson_k_1.html}
\end{links}

\section{Introduction}

Artificial intelligence (AI) and other advanced algorithmic systems are being created and adopted into a wider variety of tasks and settings than ever before, and thus it is critical to understand how this proliferation will positively or negatively impact people and society-at-large before undesirable consequences occur on a wide scale. A large program of research within scientific communities has been devoted to understanding these systems' possible impacts, especially with respect to human values such as fairness, transparency, privacy, and trust \citep{rismani2025measuring}, as evidenced by the growth of communities such as AIES and FAccT within the last 10 years \citep{VarshneyAIES, facct2025}. Following widespread trends across scientific disciplines \citep{bornmann2014altmetrics, roberts2009realizing}, the impact of publications from these venues now likely expands beyond the scientific community to policy makers, organizations, and the general public who use the research to inform their own decision making. Therefore, it is more essential than ever to ensure that scholarly publications meet the needs of people outside of the scientific community; this is especially true for scholarship engaging with potential broad societal impacts of AI systems, which inherently implicates stakeholders beyond researchers and scientists. 

Although the field of AI impact evaluation and analysis is relatively new, analyzing the potential for unintended impacts or consequences is a task that has been conducted across various disciplines for many years. For example, in the 1970s, nuclear power plant experts analyzed and discussed the likelihood of catastrophic events and whether these risks could be managed (and if so, how) \citep{aven2011quantitative}. Similar questions were also being asked about phenomena such as global warming, smoking, and food safety, necessitating the development of a scientific discipline which could integrate approaches and ideas about risk in a domain-general way. Thus, risk science (``the practice [providing] the most updated and justified [knowledge, statements, and beliefs] on risk analysis") emerged in the 1980s and has been growing since \citep{aven2025history}. In this paper, we argue that the principles of risk analysis developed through risk science are also applicable to AI impact evaluation practices, and by incorporating them, the results of scientific work can be more accessible for stakeholders beyond AI researchers (e.g., other scholars, AI users, other decision makers, etc.) to use, understand, and integrate into their own practices. 

Our aims are threefold: first, we introduce readers to the field of risk science and established principles for conducting risk analysis. Second, we conduct a systematic literature review of scientific publications evaluating the fairness of AI systems\footnote{We take an expansive view of what constitutes ``AI". As noted elsewhere \citep{aicon}, there is no consensus on definitions of AI, so we include any algorithmic artifact designed to simplify or complete tasks historically done by humans.} used for tasks in the hiring and employment domain to determine to what extent the evaluation and reporting practices used adhere to the standards established in risk science. We chose this domain because of the significant amount of existing scholarship and the wide variety of stakeholders outside of the scientific community who are impacted by AI's use in hiring and employment (e.g., job seekers, employers, government regulators, etc.) and therefore would have an interest in understanding and potentially using the results of an AI evaluation conducted by the scientific community. These factors make this task and domain ideal for determining the extent to which best practices for risk analysis and communication are followed and highlighting areas for improvement. Finally, we conduct a risk analysis case study of an AI-mediated resume screening application to demonstrate the myriad of ways risk science principles can be applied to scholarly research in the AI fairness evaluation domain and the implications of various choices. These goals are summarized by the following findings and contributions:  

\begin{enumerate}
    \item Though risk is defined by potential consequences and the uncertainties associated with them, we find that only one out of 21 metrics commonly used to evaluate hiring tools' fairness is able to quantify probabilistic uncertainty. Furthermore, results of these evaluations were reported without any variability estimates 75\% of the time, obscuring an essential component of risk characterization. 
    \item By re-evaluating data examining fairness and AI resume screening from \citet{wilson2024gender} using principles from risk science, we show how risk assessments need to incorporate both assessments of consequence severity and uncertainty to accurately and transparently characterize the risk landscape. 
    \item Finally, we introduce the AI Risk Report Card, which is an initial proposal for standardizing AI evaluation result reporting by incorporating key principles from risk science in order to make them more complete, understandable, and useful for a wide variety of stakeholders. Future work should empirically validate the utility of such a reporting tool and compare it to methods of communicating facts about AI systems (and the risks they pose).
\end{enumerate}

\section{Background}

In this section, we describe the basic principles of risk science, including what risk is, how to measure and describe it, and how to communicate about it. Throughout the discussion, we offer examples of how these principles relate to current practices in AI evaluations (broadly construed, and not limited to those evaluating resume screening fairness). 

\begin{figure*}
    \includesvg[width=\textwidth]{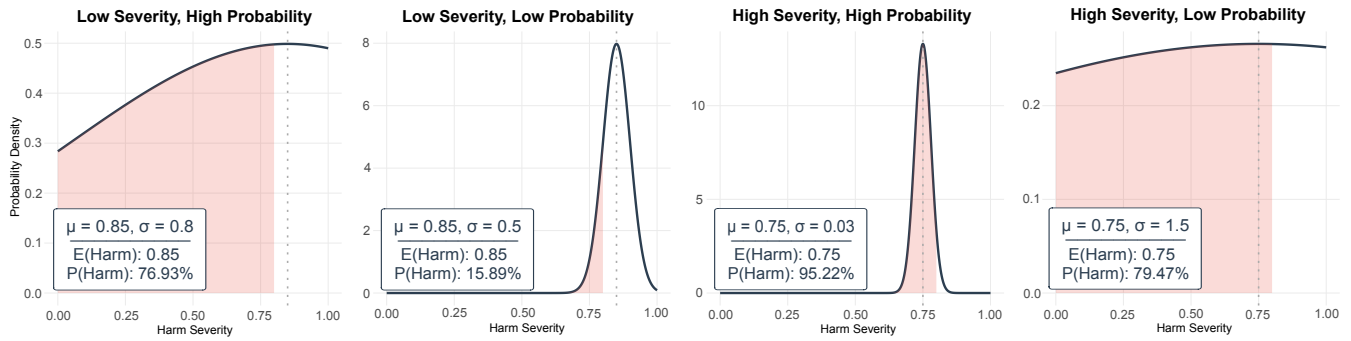}
    \caption{Illustration of how both the expected severity of a consequence and associated uncertainties play a role in risk assessment. In the two leftmost panels, the expected severity $\mu$ (0.85) is identical, but the probability that an event is harmful (any severity less than 0.8) is over 60\% greater when the variability $\sigma$ increases. If the expected severity falls within the range of harmful consequences (as in the two rightmost panels where severity is 0.75), increasing the variability has the opposite effect and decreases the probability of a severe outcome. Thus, considering both the severity and uncertainty or variability of possible consequence(s) is essential to understanding and effectively controlling or mitigating it.}
    \label{fig:illustration}
\end{figure*}

\subsection{The Meaning of ``Risk"} \label{sec:meaning_of_risk}
The use of the word ``risk" goes back hundreds of years, but as society has evolved, so has its meaning \citep{li2020brief}. In 1800, risk was strongly associated with concepts such as \textit{danger}, \textit{hazard}, and \textit{fear}. While these are still associated with risk today, there are additional stronger associations with concepts such as \textit{prevalence} and \textit{prevention} that suggest a meaning shift away from the mere representation of an event to the ability to predict, measure, or prevent it \citep{li2020brief}.

Meaning variation of risk still persists today across scientific disciplines, though at a more constrained scale than observed historical changes. For example, different definitions make different assumptions about whether risk is something that can be defined quantitatively or qualitatively; whether it exists objectively or is just derived from model concepts; if it constitutes harm alone or both harm and positive outcomes; and if it is comprised of events, uncertainties, probabilities, or some combination of these \citep{aven2012risk}. The perspective we take in this work follows from holistic definitions used in risk science--namely, that risk is a concept which exists objectively in the world, but the way that it is characterized and/or measured can be seen as inter-subjective at best \citep{aven2011ontological}. 

More specifically, risk is ``uncertainty about and severity of the consequences (or outcomes) of an activity with respect to something that humans value" \citep{aven2009risk}. To illustrate the facets of this definition in the context of AI evaluations, it is a fact independent of any individual assessor that AI model behavior cannot be perfectly predicted due to the opaque and non-deterministic nature of current state-of-the-art models, and therefore uncertainty exists about the nature of consequences that may occur when using AI models. This general existence of uncertainty reflects \textit{risk as an objective concept}. When attempts are made to characterize, assess, or attain other relevant knowledge about the unknown consequences of using AI models, then \textit{risk as a description or measurement} is dependent (i.e., subjective or inter-subjective) on the particular circumstances of its creation. This latter kind of risk is the more relevant to creators or users of AI evaluations and, unless specified otherwise, is what we mean when discussing ``risk."

Following from the definition above, the important aspects of risk to consider when attempting to measure or characterize it are 1) the consequences related to an event or activity and 2) the uncertainties associated with the event and consequences \citep{aven2011quantitative}. The consequences themselves are usually described in terms of the magnitude or severity of the impact they have on something that humans value. For example, using an untested diagnostic AI model on one patient vs. many patients could have different associated risks because the potential loss of something that humans value (e.g., health, well-being, life) is much greater when more people are exposed to the model. Since the true consequence(s) or outcome(s) cannot be known in advance of their occurrence, it is also necessary to describe the uncertainties that exist surrounding potential consequences to fully characterize the risk. Figure \ref{fig:illustration} shows an illustration of this based on disparate impact in employment, which states that an unfair outcome occurs if any group is selected at less than 80\% of the rate of the group with the highest selection rate \citep{fabris2025fairness}. Even when the average severity of disparate impact stays constant, changing the uncertainty surrounding the expected harm of the outcome can vastly change the overall risk associated with a particular scenario.  

The differences illustrated in Figure \ref{fig:illustration} reflect \textit{aleatory}/\textit{stochastic} uncertainties, which are variations or randomness among units that are similar to the one being studied but not observed directly \citep{aven2011different}. This kind of uncertainty is likely familiar to many quantitative researchers who use inferential statistics methodologies, in which probability theory is used to draw conclusions about the most likely characteristics of a population using only observations from a small sample \citep{thekdi2024understanding}. Another type of uncertainty that is necessary to characterize risk is \textit{epistemic uncertainty}, which reflects a more general lack of knowledge about phenomena or the world and its processes \citep{aven2014uncertainty}. Crucially, epistemic uncertainty can be reduced or eliminated by gaining more knowledge about the world, while stochastic uncertainty can never be fully removed. Returning to the example in Figure \ref{fig:illustration}, 0.8 was chosen as a threshold for disparate impact unfairness based on legal precedent and tradition \citep{fabris2025fairness}, but there is no empirical evidence (currently) that this threshold meaningfully represents unfairness better than an alternative (e.g., 0.7 or 0.9) might. 

The lack of knowledge about what the ``best" choices to model unfairness in the employment context reflects epistemic uncertainties that naturally exist across every domain and task. However, characterizing these uncertainties in any risk assessment is necessary to interpreting its results---for example, if epistemic uncertainty is not specified, then decision makers might make ill-informed choices based on information with only weak support \citep{thekdi2024classification}. Alternatively, if epistemic certainty is known, it is clear whether uncertainty in the risk assessment can be decreased through further information gathering and analysis or if it is entirely stochastic and irreducible \citep{aven2011some}.

\subsection{Assessing Consequences and Uncertainties} \label{sec:assess_consequences}

The first step of assessing risk for a given activity is identifying potential credible consequences or outcomes that could occur, usually with the involvement of domain experts. The initial list of possibilities may include outcomes that have no record of previously occurring, but these should still be included to ensure that consequences which are rare but high severity are not overlooked \citep{aven2015implications, ostrom2019risk}. Once a set of possibilities is identified, the most important (with respect to causing the greatest loss or harm to things people value) can be characterized and analyzed further. In this analysis, assessors may look to historical data or testable models of systems to determine the possible ranges of consequence severities or magnitudes. In AI evaluation, this could mean reviewing reports of past harmful events archived in places like the AI Risk Repository \citep{slattery2026ai} or modeling the usage of AI models in the real world through paradigms like benchmarking, red-teaming, or auditing \citep{solaiman2025}. Throughout all of these, descriptive statistics (e.g., identifying the range of possible consequence magnitudes and severities) are important to characterizing consequences quantitatively. 

Once the severity and/or magnitude of consequences has been identified descriptively, it is necessary to identify the most likely outcomes and the level of (un)certainty about this judgment (i.e., stochastic uncertainty). As mentioned above, statistical inference can play a critical role here, as it enables assessors to use the distribution of outcome possibilities to draw conclusions about what values unobserved samples would be likely to take \citep{aven2011quantitative}. For example, if a researcher was interested in whether an AI system is able to accurately diagnose a disease, they might use the values observed in a random sample of correct and incorrect AI diagnoses to determine what the theoretical expected value (i.e, center of mass in a probability distribution) of accuracy is, how much this value could vary in unobserved samples using confidence intervals, or how likely it is that the system performs better than chance or experts using null hypothesis significance testing. All of these quantities can be derived using only frequencies (and associated frequentist probabilities) from sample observations \citep{aven2011quantitative}.

However, frequentist probabilities are insufficient to describe epistemic uncertainty, and risk scientists often suggest using \textit{subjective probabilities} instead \citep{aven2011interpretations}. Under these frameworks, probability is an expression of uncertainty as perceived by an assessor using evidence such as statistical assessments, direct experience, models, or other theoretical approximations \citep{aven2009risk}. One of the most well known subjective probability paradigms is Bayesian probability, which relies on an assessor's prior knowledge of an event to characterize its probability; when the assessor receives new information, their priors are updated to accommodate this into the probability model \citep{aven2020bayesian}. Other methods that have been suggested include imprecise or interval probabilities, fuzzy probabilities, possibility theory, and evidence theory \citep{aven2011interpretations}. 

For simplicity, we will discuss epistemic uncertainty using interval probabilities, which are statements made using ranges of possible probabilities rather than a single value. For instance, rather than saying a particular AI model has a 40\% chance of producing inaccurate information, an evaluator could quantify their epistemic uncertainty by saying the chance is between 25\% and 50\%. The interpretation of such a subjective probability statement is not the same as typical frequentist probability, but rather should be interpreted as a statement about an assessor's beliefs given the knowledge available to them. It is often helpful to compare these subjective probabilities to reference events which have commonly accepted interpretations, such as flipping coins or rolling dice \citep{aven2013define}. In this instance, the assessor is saying that they believe the uncertainty around inaccurate information being produced is at least the uncertainty of getting two heads on two consecutive coin flips and at most the uncertainty of getting one head in one flip.

Finally, any uncertainty estimations should also be qualified with descriptions of the knowledge used to derive them and its associated strength. The aspects of knowledge that are relevant are the same as those which would be included in any other scientific inquiry--for example, methods and materials used to conduct any empirical investigations, data analysis procedures, key assumptions, related information from outside sources, and theoretical frameworks, among many others. The choices made to generate knowledge can then be qualitatively categorized as producing strong, moderate, or weak knowledge, using the criteria proposed by \citet{flage2009expressing} (reproduced in the appendix in the extended version of the paper, for example, or other similar qualitative assessment tools (see \citet{askeland2017moving}). Importantly, the strength of knowledge is primarily a tool for decision makers or other stakeholders outside of the assessment team to contextualize and apply the results of the risk assessment for their needs (e.g., whether to gather more data, make an investment, proceed with or change a proposed plan, etc.), and it is not a reflection of the overall quality of the risk assessment \citep{aven2011misconceptions}. A risk assessment which uses sound methods to identify potential consequences and associated uncertainties but has limited or unreliable data to examine would be categorized as weak by \citet{askeland2017moving}'s criteria, but its quality could be considerably higher than one who has more reliable data available but is less transparent about the assumptions and methods used. Ultimately, the knowledge about risk and the way it is used to derive consequences and their associated uncertainties can be judged as any other scientific activities would be: through the lens of validity (i.e., how faithfully the real-world phenomena of interest are represented) and reliability (i.e., how consistent results are when the investigation is repeated) \citep{aven2009reliability}.

\subsection{Perceiving and Communicating Risk}

The final step of risk analysis for the assessor is to communicate the findings and results to people who were not part of the assessment team. While identifying the best ways of communicating risk is still an open research area \citep{aven2025uncertainty}, there is some consensus that communicating probabilities shared by strength of knowledge judgments is preferable to sharing probabilities with other kinds of uncertainty judgments because they are more informative, trustworthy, and justifiable \citep{thekdi2025evaluating}. However, despite experts and risk analysts' best efforts to communicate risks accurately and faithfully, lay people (including those who might be ultimately responsible for using the risk assessments to make decisions) may have different understandings of the risks presented. For example, people may perceive something to be more or less risky based on how familiar they are with the particular circumstances under consideration or how much control they believe they have over the circumstances or consequences under consideration \citep{aven2019science}. Ultimately, further research into people's risk literacy will be necessary to establish additional best practices beyond those mentioned here \citep{aven2024risk}.

The necessity of high-quality communication of AI evaluation results is also an active area of research. Inspired by model and data cards \citep{mitchell2019model, pushkarna2022data}, which aim to succinctly summarize important information about the development and intended uses of opaque technological artifacts for stakeholders outside of development teams, multiple variations of AI impact cards have been proposed for similar purposes; risk communication manifests differently in each of these. For example, \citep{gursoy2022system}, \citep{staufer2025audit}, \citet{bogucka2025impact} and state that including information about risk on their AI impact cards is important, but they provide very little information about how to describe various dimensions of risks or how risk communication differs from communicating other aspects like model capabilities. \citet{golpayegani2024ai} design AI impact cards specifically based on the EU AI Act, which explicates possible risks associated with AI use in detail \citep{novelli2024taking}. They include information about risk likelihood and severity, but provide little information about associated uncertainties that are essential for risk analysis and communication, as discussed in Section \ref{sec:assess_consequences}. Future work is needed at the intersection of risk and AI literacy in order to achieve high-quality risk communication in this domain.

\section{Risk in AI Hiring Fairness Evaluations}

In this section, we describe a systematic analysis of metrics used to evaluate AI tools used for hiring or employment tasks to determine the extent to which the principles of risk science are already incorporated into AI evaluation practices. We focus on evaluations of model fairness conducted primarily by and for the scientific community (i.e., those published at scholarly venues rather than as part of business or journalistic reporting). The results of this survey can elucidate which aspects of risk researchers are already measuring and describing as part of their scientific inquiries, which aspects are currently underrepresented in published scholarship and, accordingly, how this impacts people outside of the scientific community's ability to use scientific inquiry as part of their risk assessment and decision making.

\subsection{Methods}

We began our review of metrics by annotating the metrics listed in \citet{fabris2025fairness}, a multidisciplinary survey of fairness and bias in algorithmic hiring. By conducting a systematic literature review, they identify 21 unique metrics which are used for fairness evaluation across 22 different studies (on average each metric appears in 1.66 studies as some are used in multiple studies). \citet{fabris2025fairness} categorize each metric according to what kind of fairness it relates to (\textit{outcome, impact, accuracy, representational, impact,} or \textit{process}). They also annotate features of the metrics related to their interpretation and use, such as whether they allow for the inclusion of multinary social or demographic attributes or if they account for potential differences in user hyperparameter choices. We expand on these features by adding annotations for features related to risk assessment, such as whether they characterize aspects of severity or stochastic uncertainty. 

For each metric, we first determine what the mathematical bounds or limits are for the metric. For example, a metric like disparate impact (which is typically formulated as the ratio between selection rates of the minority and majority groups) could have a minimum value of zero when the minority/majority group is never/always selected or a maximum value of one if the minority and majority groups have the same selection rates. Next, these bounds were used to annotate whether each metric satisfied the constraint that the probability $P$ of any subset of events $A$ in the set of all possible events $F$ must be at least zero and at most one (Equation \ref{eq:bounds}) \citep{sep-probability-interpret}. For metrics that satisfied this constraint, we annotated whether they satisfied the second constraint of probability: additivity (Equation \ref{eq:additivity}) \citep{sep-probability-interpret}. This criteria states that the probabilities of all possible outcomes for a given event should sum to one exactly. Metrics which satisfied both bounds and additivity constraints were thus categorized as probability metrics (which can be used to refer to the uncertainty of a consequence), and those which did not satisfy both the constraints were categorized as severity metrics (which can characterize potential consequences). 

Second, we reviewed each of the studies cited by \citet{fabris2025fairness} and annotated whether their experimental results using the surveyed metrics were reported as point estimates only or as point estimates with variability estimates (e.g., confidence intervals, standard deviations, distributions, etc.). If variability estimates were included, we noted which type was used and whether these estimates were given for all results or only a subset. As no instances of formalizing epistemic uncertainties in AI evaluations were found in an initial review of the studies listed by \citet{fabris2025fairness}, we did not include identifying or describing this type of uncertainty in our annotation procedure. 

\begin{equation}
    0 \leq P(A) \leq 1 \quad \forall A \in F
    \label{eq:bounds}
\end{equation}
\begin{equation}
    P\left(\bigcup_{i=1}^{\infty} A_i\right) = \sum_{i=1}^{\infty} P(A_i)
    \label{eq:additivity}
\end{equation}

\subsection{Results and Discussion}

The results of our metric review and annotation are shown in Table \ref{tab:severity}. In the remainder of this section, we discuss the patterns we identified and their implications for AI fairness risk assessment and communication.

\begin{table}[]
\centering
\setlength{\tabcolsep}{4pt} %
\renewcommand{\arraystretch}{1} %
    \begin{tabular}{@{}lcc@{}}
        \toprule
        \textbf{Measure} & {\color[HTML]{333333} \textbf{Bounds}} & {\color[HTML]{333333} \textbf{\begin{tabular}[c]{@{}c@{}}Valid \\ Probability\end{tabular}}} \\ \midrule
        \rowcolor[HTML]{D9EAD3} 
        Skew@k & {\color[HTML]{333333} (-$\infty$, $\infty$)} & {\color[HTML]{333333} No (I)} \\ [.1cm] 
        \rowcolor[HTML]{D9EAD3} 
        \begin{tabular}[c]{@{}l@{}}Normalized Discounted\\Cumulative Kullback-Leibler \\ Divergence (NDKL)\end{tabular} & {\color[HTML]{333333} {[}0,1{]}} & {\color[HTML]{333333} No (II)} \\ [.5cm] 
        \rowcolor[HTML]{D9EAD3} 
        Disparate Impact (DI) & {\color[HTML]{333333} {[}0,1{]}} & {\color[HTML]{333333} No (II)} \\ [.1cm] 
        \rowcolor[HTML]{D9EAD3} 
        Demographic Disparity (DD) & {\color[HTML]{333333} {[}-1,1{]}} & {\color[HTML]{333333} No (I)} \\ [.1cm] 
        \rowcolor[HTML]{D9EAD3} 
        \begin{tabular}[c]{@{}l@{}}Representation in Positive \\ Predictive Rate (RPPR)\end{tabular} & {\color[HTML]{333333} {[}-1,1{]}} & {\color[HTML]{333333} No (I)} \\ [.3cm] 
        \rowcolor[HTML]{D9EAD3} 
        \begin{tabular}[c]{@{}l@{}}True-Positive Rate \\ Difference (TPRD)\end{tabular} & {\color[HTML]{333333} {[}-1,1{]}} & {\color[HTML]{333333} No (I)} \\ [.3cm] 
        \rowcolor[HTML]{D9EAD3} 
        False-Negative Rate Ratio (FNRR) & {\color[HTML]{333333} {[}0,$\infty$)} & {\color[HTML]{333333} No (I)} \\ [.1cm] 
        \rowcolor[HTML]{D9EAD3} 
        \begin{tabular}[c]{@{}l@{}}Extended Equality of \\ Opportunity (ExEO)\end{tabular} & {\color[HTML]{333333} {[}0,1{]}} & {\color[HTML]{333333} No (II)} \\ [.3cm] 
        \rowcolor[HTML]{D9EAD3} 
        \begin{tabular}[c]{@{}l@{}}Discounted Representation \\ Difference (DRD)\end{tabular} & {\color[HTML]{333333} {[}-1,1{]}} & {\color[HTML]{333333} No (I)} \\ [.3cm] 
        \rowcolor[HTML]{D9EAD3} 
        Score KL Divergence (SKLD) & {\color[HTML]{333333} {[}0,$\infty$)} & {\color[HTML]{333333} No (I)} \\ [.1cm] 
        \rowcolor[HTML]{D9EAD3} 
        Log Rank Regression (LRR) & {\color[HTML]{333333} (-$\infty$, $\infty$)} & {\color[HTML]{333333} No (I)} \\ [.1cm] 
        \rowcolor[HTML]{D9EAD3} 
        \begin{tabular}[c]{@{}l@{}}Root Mean Square of TPRD\\(RMS)\end{tabular} & {\color[HTML]{333333} [0,1]} & {\color[HTML]{333333} No (II)} \\ [.3cm] 
        \rowcolor[HTML]{D9EAD3} 
        Mean Error Difference (MED) & {\color[HTML]{333333} (-$\infty$, $\infty$)} & {\color[HTML]{333333} No (I)} \\ \midrule
        \rowcolor[HTML]{C9DAF8} 
        Mean Absolute Error (MAE) & {\color[HTML]{333333} (-$\infty$, $\infty$)} & {\color[HTML]{333333} No (I)} \\  [.1cm] 
        \rowcolor[HTML]{C9DAF8} 
        \begin{tabular}[c]{@{}l@{}}Balanced Classification \\ Rate Difference (BCRD)\end{tabular} & {\color[HTML]{333333} {[}-1,1{]}} & {\color[HTML]{333333} No (I)} \\ [.3cm]
        \rowcolor[HTML]{C9DAF8} 
        \begin{tabular}[c]{@{}l@{}}Mutual Information Difference \\ (MID)\end{tabular} & {\color[HTML]{333333} {[}-1,1{]}} & {\color[HTML]{333333} No (I)} \\
        \rowcolor[HTML]{D9D2E9}  \midrule
        Salary Difference (SaD)& {\color[HTML]{333333} (-$\infty$, $\infty$)} & {\color[HTML]{333333} No (I)} \\ \midrule
        \rowcolor[HTML]{F4CCCC} 
        Gender Bias Score (GBS)& {\color[HTML]{333333} (-$\infty$, $\infty$)} & {\color[HTML]{333333} No (I)} \\ \midrule
        \rowcolor[HTML]{FFF2CC} 
        \begin{tabular}[c]{@{}l@{}}Sensitive Area Under the \\ Receiver Operating \\ Characteristic Curve (sAUC)\end{tabular} & {\color[HTML]{333333} {[}0,1{]}} & {\color[HTML]{333333} Yes} \\ [.5cm]
        \rowcolor[HTML]{FFF2CC} 
        Ground Truth Regression (GTR)& {\color[HTML]{333333} (-$\infty$, $\infty$)} & {\color[HTML]{333333} No (I)} \\ [.15cm] 
        \rowcolor[HTML]{FFF2CC} 
        \begin{tabular}[c]{@{}l@{}}Mutual Information \\ Amplification (MIA)\end{tabular} & {\color[HTML]{333333} {[}-1,1{]}} & {\color[HTML]{333333} No (I)} \\ \bottomrule
    \end{tabular}
    \caption{Summary of whether the AI hiring fairness metrics listed in \citet{fabris2025fairness} describe severity or uncertainty of harms. If a metric is not a valid probability (all but one), the constraints it violates are given in parentheses (corresponding to criteria in Equations \ref{eq:bounds} and \ref{eq:additivity}). The colors of rows correspond to the different kinds of fairness identified by \citet{fabris2025fairness}: \colorbox{MyGreen}{Outcome}, \colorbox{MyBlue}{Accuracy}, \colorbox{MyPurple}{Impact}, \colorbox{MyRed}{Representational}, and \colorbox{MyYellow}{Process}.}
    \label{tab:severity}
\end{table}

\subsubsection{Metrics Are Dominated by Severity} As shown in Table \ref{tab:severity}, the vast majority of metrics (20/21) used to assess the fairness of AI hiring tools are not valid probabilities, and therefore characterize the severity of the unfairness rather than uncertainty about whether it occurs.\footnote{Equations for each metric as well as the list of studies each was used in is shown in the appendix.} The only exception to this pattern is the Sensitive Area Under the Receiver Operating Characteristic Curve (sAUC). The sAUC is a specific application of the more general Area Under the Receiver Operating Characteristic Curve (AUC-ROC), which is a function of True Positive Rates (TPR) and False Positive Rates (FPR) at various classification thresholds. As TPR and FPR can both take values from zero to one and are themselves valid probabilities, the area under any curve produced by plotting TPR and FPR against each other is also between zero and one and a valid probability \citep{hanley1982meaning}. 

In the case of sAUC, the classification task is to predict a sensitive attribute (e.g., gender) using non-sensitive features that are also used in a prediction task related to hiring. The ideal outcome in this scenario is that sAUC is approximately 0.5, meaning that the classifier's input features only afford classification at chance levels (e.g., the input features contain no information about the sensitive attributes). If sAUC is higher, it implies that there is an equivalent increase in the probability that a sensitive attribute could be correctly predicted from the input features. Therefore, the interpretation of sAUC in the AI hiring risk assessment context is that it describes the uncertainty of whether or not a person's social or demographic identities are recoverable from other ``non-sensitive" data that are used in the AI task. Other metrics would be needed to describe the severity of this consequence, such as how much or in what way this recoverability impacts the outcome of hiring or employment decisions. 

The dominance of severity measures implies that important aspects of risk related to the use of AI hiring and employment tools have been overlooked in AI fairness evaluations up to this point. For example, while it may be important to know the expected level of disparate impact when using an AI hiring tool with no other information available, this does not provide information about how likely disparate impact (of any magnitude) would be to occur in a given situation. Both pieces are important to understanding the risk of a system in order to make decisions about it, as strategies could differ significantly based on the chance of a consequence occurring independent of its expected severity. 

\begin{figure}
    \includesvg[width=0.4\textwidth]{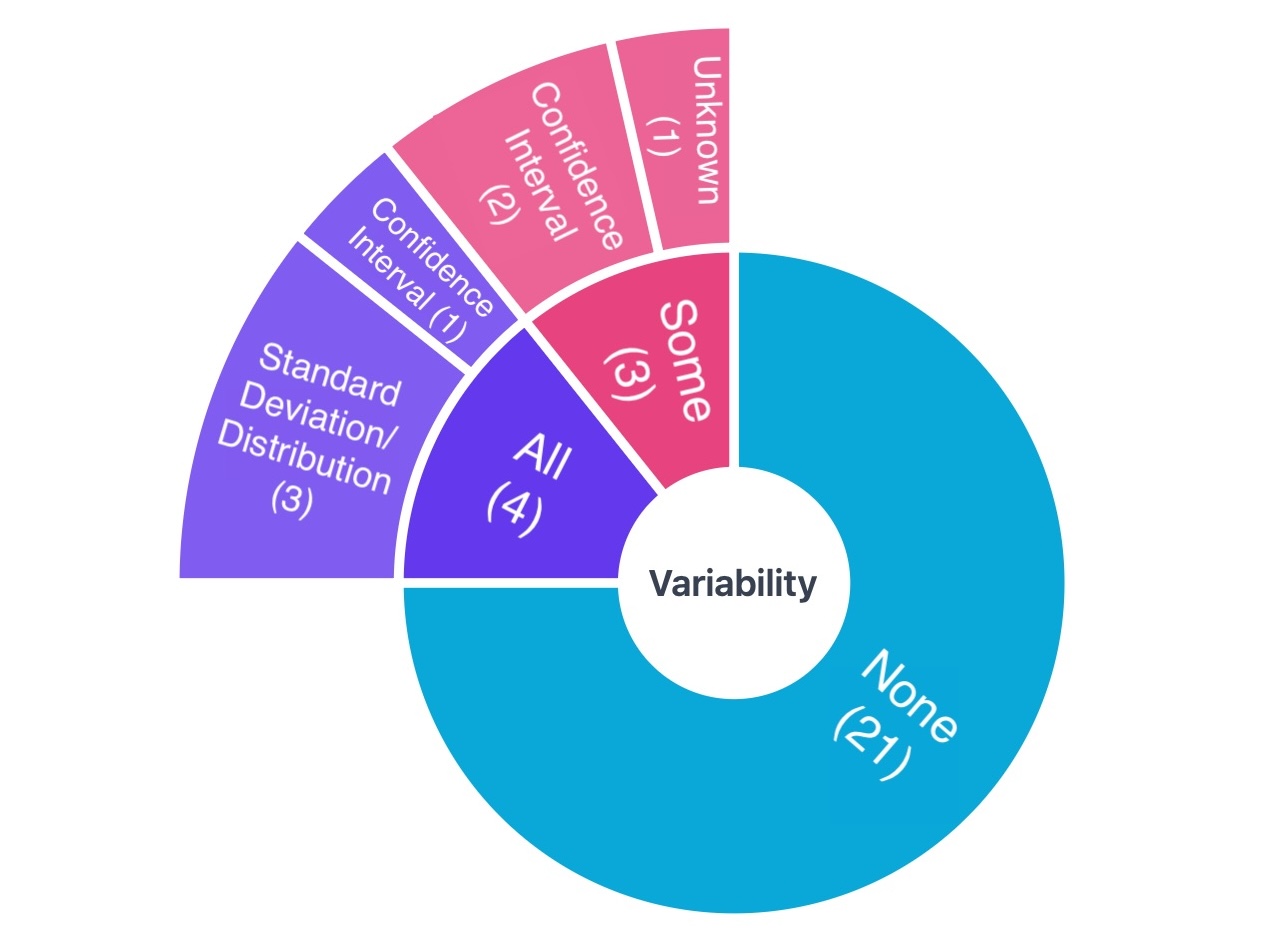}
    \caption{Among the 28 evaluations in our analysis set, only seven provided variability estimates for their metric results. Of those, four provided variability estimates for all results: one used a confidence interval, and three used a combination of standard deviations or errors and visualizations of distributions. Two papers provided confidence intervals for some but not all results, and one provided variability estimates for some results but did not specify the type.}
    \label{fig:sunburst}
\end{figure}

\subsubsection{Results Lack Associated Variability Estimations} As shown in Figure \ref{fig:sunburst}, across the 28 reports of results using the metrics listed in Table \ref{tab:severity}, 21 of them (75\%) are given as point estimates only, without any associated measure of variability. In four examples, every time a result was given using the metrics in Table \ref{tab:severity}, it was accompanied by some form of variability estimate: one of these used confidence intervals and the other three used a combination of standard deviations or errors and visualizations of the distribution of results. In general, the visualization of full distributions is the most informative about uncertainty because they are able to communicate the overall range and shape of data as well as measures of central tendency like the mean, median, or mode. Confidence intervals and standard deviations or errors provide much less detailed information and can be difficult to interpret depending on the distributional characteristics of the observations used to produce the measures. 

\subsubsection{Communicating Epistemic Uncertainty is Challenging.} 
Reflecting trends in other disciplines, there is no reporting standard to communicating epistemic uncertainties \citep{zehr2012scientists, roney2026bread}, which we observed in our initial review of the 22 hiring evaluation studies. As epistemic uncertainties are a critical part of risk assessment, high-quality AI evaluations should have clear discussions of what epistemic uncertainties exist and specific details about how they could impact the interpretation of assessed risks.

\section{Case Study: Evaluating Fairness Risk of AI Hiring Tools Using Risk Science Principles}
In this section, we use AI-mediated resume screening as a case study to demonstrate how the principles of risk science can be incorporated into AI fairness evaluations to improve the clarity and utility of scholarly work for stakeholders both inside and outside of the scientific community. We start by describing and discussing a previously published evaluation of fairness in AI resume screening and conclude by re-analyzing and reporting the data to better adhere to the risk science principles established previously.

\subsection{Summary of \citet{wilson2024gender}}

\subsubsection{Resume Screening Approach and Data}
Before risk analysis can begin, it is necessary to have data to inform the assessments. Though many sources of data could be used depending on the goals, for this case study we chose to use data that closely matches what researchers use in typical AI fairness evaluations (e.g., outputs or behaviors elicited from AI models in experimental or observational settings) to demonstrate how methods which are already commonly used and accepted by the scientific community can be adapted to be more accessible and informative to those outside of it. Specifically, we conduct this case study using data from \citet{wilson2024gender}, a study which investigated intersectional (gender and race) biases in resume screening conducted by large language models (LLMs).

\citet{wilson2024gender} used LLMs to rank hypothetical candidate resumes for a variety of open job positions. Each position was represented by a real job description that had been posted online, and resumes were also based on real resumes, though the researchers added 80 names signaling Black men, Black women, white men, and white women to the anonymous resumes to create a set of documents that represented candidates with different social identities who were equally qualified for each job. They evaluated the fairness of three LLMs (E5-mistral-7b-instruct (e5); GritLM-7B (GritLM); SFR-Embedding-Mistral (SFR)) used to rank the candidates for the job descriptions representing nine different occupations (Chief Executive, Marketing and Sales Manager, Miscellaneous Manager, Human Resource Worker, Accountant and Auditor, Miscellaneous Engineer, Secondary School Teacher, Designer, and Miscellaneous Sales and Related Worker). In total, there were approximately 9 million data points to use for fairness evaluation (comprised of 39,814 pairs of job descriptions and sets of 80 equally qualified candidates of different social identities, scored by three different LLMs). 

\subsubsection{Analysis and Results Reporting}
\begin{figure}
    \begin{tcolorbox}[colback=MyBlue,colframe=RoyalBlue,title=Risk as Reported by \citet{wilson2024gender}]
    \textbf{Consequence Severity:} In rankings of the top 10\% of candidates, resumes with white names have an $\approx$10\% advantage over those with Black names when screening Accountants and Auditors with GritLM. 
    \tcblower
    \textbf{Consequence Uncertainty:} In rankings of the top 10\% of candidates, resumes with white or Black names are significantly preferred (p$<$0.05) in 85.1\% or 8.6\% of tests, respectively.
    \end{tcolorbox}
    \caption{Examples of risk characterizations given in the AI resume screening evaluation conducted by \citet{wilson2024gender}.}
    \label{fig:results2024}
\end{figure}

\citet{wilson2024gender} reported the results of their LLM evaluation at the level of model and/or occupations by averaging the fairness risk metrics computed for each combination of job description and corresponding candidate ranking. Two examples of such results are shown in Figure \ref{fig:results2024}. The first result describes the severity of LLM resume screening consequences using a metric akin to Demographic Disparity (DD). Because the difference in selection rates is not itself a valid probability, this metric alone does not characterize any stochastic consequence uncertainty. The second result, however, does characterize uncertainty as it is based on the frequency of rankings with statistically significant disparities relative to the total number of rankings. Such a metric satisfies both the criteria in Equations \ref{eq:bounds} and \ref{eq:additivity}, and thus can characterize the stochastic uncertainty of LLM resume screening. 

Each of the results presented in Figure \ref{fig:results2024} does not fully characterize unfairness risks associated with LLM resume screening. In order for the consequence severity result to be more informative of overall risk, the stochastic uncertainty of the severity should also be assessed. This would give evidence as to whether the most likely outcome is very likely (e.g., is expected to happen 90\% of the time, and other possible severities have a much lower chance of occurring) or very unlikely (e.g., is expected to happen only 10\% of the time, and other possible severities have only slightly smaller chances of occurring). The implications for stakeholders beyond the AI evaluators could be very different depending on how certain the prediction of the expected severity is.

In contrast, the second result in Figure \ref{fig:results2024} does characterize the stochastic uncertainty of the consequences, but it does not describe consequence severity. The statement that resumes with white names are significantly preferred indicates that a hypothesis test was conducted, and the results suggest with a high level of certainty that differences between rankings of resumes with white vs. Black names can likely be attributed to underlying differences in the way they are evaluated by the LLM and not chance (i.e., the existence of racial bias). Furthermore, these significant differences were observed in the majority of examples (85.1\%). Therefore, the interpretation of this result in a predictive risk sense is that if given a set of equally qualified white and Black candidates, the chance that the white candidates would be ranked more highly than Black candidates is 85.1\%. This interpretation does not give any information about how severe this ranking disparity is likely to be though---it could be relatively modest (e.g., the top 50 spots are occupied by 30 white candidates and 20 Black candidates) or extremely severe (e.g., the top 50 spots are occupied by all white candidates and no Black candidates). As above, the implications for stakeholders beyond the AI evaluators could also be very different depending on which level of severity is more likely. 

Finally, in neither result in Figure \ref{fig:results2024} is epistemic uncertainty formalized or used to further qualify the risk estimates. This adds additional challenges to interpreting and utilizing \citet{wilson2024gender}'s findings, as it is unclear what different circumstances (if any) might lead to a significantly altered risk landscape. Due to the limitations identified here, the way that \citet{wilson2024gender} has presented their results can be significantly improved by incorporating additional characterizations of consequence severity, stochastic uncertainty, and epistemic uncertainty, in order to create a fuller assessment of the risk associated with using LLMs for resume screening.

\subsection{Using Risk Science Principles to Re-Analyze \citet{wilson2024gender}}

\subsubsection{Describing Consequence Severity} 

Just as \citet{wilson2024gender} did, our goal for this analysis is to determine the risk of using LLMs for resume screening in relation to selecting candidates which are best qualified for roles and making selections based on social identities such as race or gender. In other words, the consequence we are interested in analyzing is the selection of  unequal numbers of equally qualified candidates for a given job based on their social identities, and (some of) the value(s) relevant to this consequence are justice, equality, and non-discrimination. 

Using the same data as \citet{wilson2024gender}, we first analyze the consequence of interest descriptively by calculating the Disparate Impact (DI), which is the ratio of the lowest and highest selection rates when selecting $k$ candidates among all groups $g$ in the set of groups $S$ (Equation \ref{eq:DI}). We chose to select the top 50 candidates as an initial point of analysis. As noted previously, the ``80\% rule" is a heuristic to differentiate discriminatory DI ($DI < 0.8$) from non-discriminatory ($DI \geq 0.8$), and in practice an LLM is considered fair if most scores are between 0.8 and 1; values near zero indicate high unfairness or discrimination. 

\begin{align}
    \text{DI} &= \frac{\text{min}_{g \in S} N_g^k / N_g}{\text{max}_{g \in S}N_{{g}}^k / N_{{g}}} \label{eq:DI}
\end{align}

\begin{figure*}
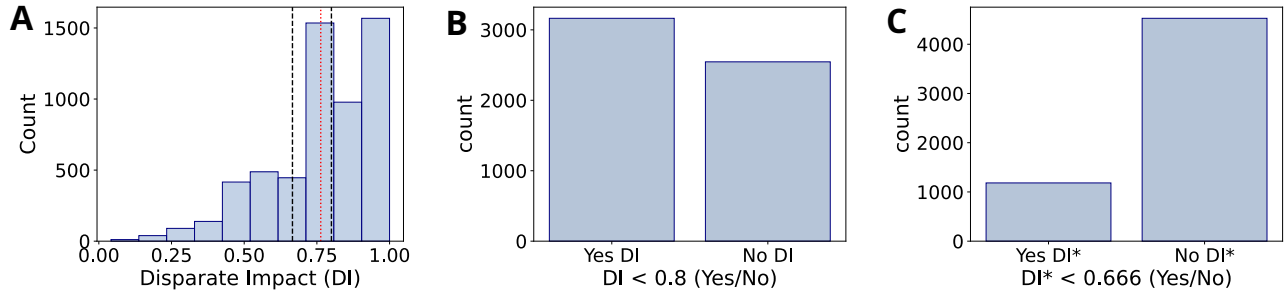

    \centering
    {\includesvg[width=0.32\textwidth]{Images/case_DI_distribution_GritLM.svg}}
    {\includesvg[width=0.32\textwidth]{Images/case_DI_YN_distribution_GritLM.svg}} 
    {\includesvg[width=0.32\textwidth]{Images/case_DIsig_YN_distribution_GritLM.svg}}
    \caption{[A] The distribution of observed Disparate Impact (DI) scores when evaluating white men's vs. Black men's resumes using GritLM. The red line marks the sample average; the two dashed black lines represent different possible DI thresholds at 0.666 and 0.800. [B] The number of resume screening scenarios that are classified as discriminatory vs. non-discriminatory when the DI threshold is 0.800 (based on the 80\% rule). [C] The number of resume screening scenarios which are classified as discriminatory vs. not when the DI threshold is 0.666 (based on statistical significance).}
    \label{fig:case_results}
\end{figure*}

As shown in Figure \ref{fig:case_results} for the model GritLM, the average disparate impact observed across all scenarios is 0.763 (which is below the 0.8 threshold indicating that unfairness has occurred).\footnote{\citep{wilson2024gender} evaluated two additional models which are omitted here for brevity. The other models showed similar patterns as illustrated in Figure \ref{fig:case_results}, and these supplementary results are available in the appendix of the extended version of the paper.} However, the spread of observed values spans nearly the total range of possible DI values, with the minimum DI at 0.042 and the maximum at 1.0. This indicates that there is some amount of variability in the values that could be expected on average in new samples from the same population of job descriptions and resumes. Furthermore, the distribution is left-skewed, suggesting that other descriptive measures to characterize variance (such as standard deviation) would not be appropriate in this case.   

\subsubsection{Describing Stochastic Consequence Uncertainty} 

First, to determine what range of harm is most likely to occur on average, we conducted 5,000 iterations of bootstrap sampling to construct a 95\% confidence interval showing that the average value will fall between 0.759 and 0.768 in most of the similar samples. This suggests that there is a high degree of (stochastic) certainty that the harm that is most likely is severe enough to be considered discrimination, but not the highest magnitude of discrimination possible. 

While looking at the range of expected average DI values can inform stakeholders that harm is more likely to occur than not, it does not provide specific information about that likelihood. To analyze this, we code the instances where DI occurred (i.e., was less than 0.8) or not as a binary factor and sum the frequencies of each level, which is shown in Figure \ref{fig:case_results}. As expected, discriminatory DI outcomes occur more often than non-discriminatory outcomes (55.41\% of the time). We follow the same bootstrap sampling procedure done for raw DI values for the recoded binary values and find that the 95\% confidence interval is again quite narrow, with most similar samples having harm occur in 54.11-56.70\% of resume screening scenarios. 

\subsubsection{Describing Epistemic Uncertainty} As demonstrated above, this evaluation has a relatively small amount of stochastic uncertainty, and so any remaining uncertainty is attributable to a lack of knowledge about the processes, applications, and scale of resume screening and discrimination in the real world. For example, as discussed in Section \ref{sec:meaning_of_risk}, the 80\% rule is not empirically validated nor the only standard that can be used to determine whether or not discrimination occurs \citep{watkins2024four, meier1984happened}. In Figure \ref{fig:case_results}, we show that using a different common threshold based on statistical significance\footnote{According to the chi-square distribution, the smallest disparity that is still significant at the $\alpha=0.05$ level with 50 candidates from two different groups is 20 vs. 30. Therefore the DI threshold based on statistical significance is $20 / 30 = 0.666$.} can lead to very different risk characterizations. 

As changing the threshold at which a consequence is considered harmful discrimination does not change the distribution of data itself, the expected severity values remain the same as when evaluating DI at the 0.8 threshold and are still most likely between 0.759 and 0.768. However, the threshold does change the interpretation of this value---at the lower threshold of 0.666, most of the observed outcomes would not be considered harmful discrimination. Therefore ,the likelihood of harm occurring decreases drastically and occurs roughly 20.72\% of the time (95\% CI: 19.67-21.75\%). This change in predicted likelihood demonstrates how characterizing epistemic uncertainty can be useful for stakeholders who may want to gather more information to reduce epistemic uncertainty before making decisions or otherwise acting on the risk assessment results. Accordingly, a possible subjective probability description of AI resume screening using interval probabilities could be that harm has at least a 19.67\% and at most a 56.70\% chance of occurrence.

In practice, describing every possible source of epistemic uncertainty and associated changes in risk beliefs is likely impossible. For example, it would be very difficult to assess every possible DI threshold; further epistemic uncertainty also exists for whether DI is the best metric to assess fairness. Just determining the bounds of risk for all combinations of thresholds and fairness metrics is unfeasible, and yet there are still many other sources of epistemic uncertainty that could be incorporated. Thus, most current recommendations in risk science suggest characterizing epistemic uncertainties using broad qualitative judgments based on loose criteria about what constitutes ``weak knowledge" vs. ``strong knowledge" \citet{askeland2017moving}. For example, in this analysis, knowledge would be characterized as weak since the risk assessment involved strong simplifications and assumptions about how AI-mediated resume screening works in the real world (e.g., the LLMs used, the extent of variability in candidate resumes, the lack of human interaction or review, etc.). The full description of strength of knowledge levels and criteria is available in the appendix.

\subsubsection{Communicating Risk: The AI Risk Report Card} 
\begin{figure*}
\centering
    \includesvg[width=0.7\textwidth]{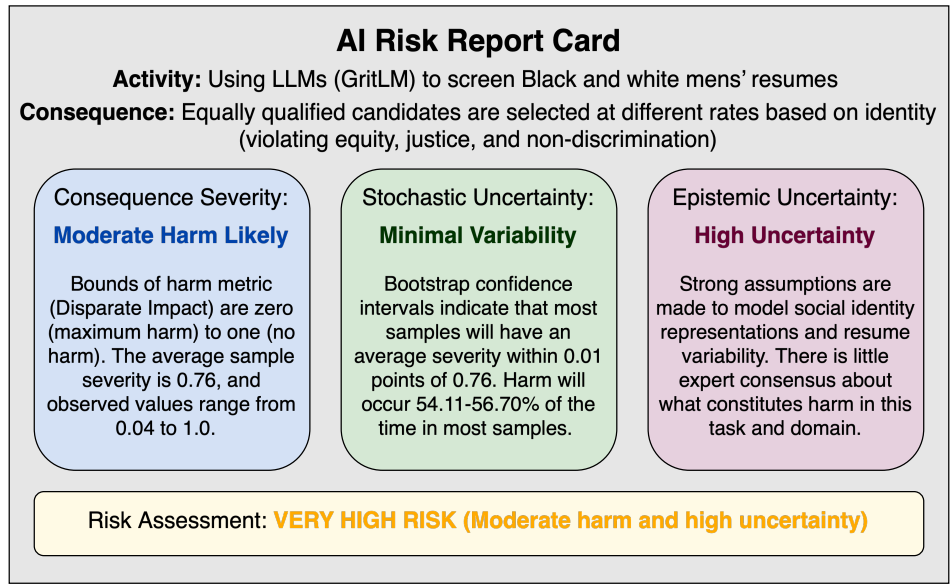}
    \caption{Example of a proposed AI Risk Report Card to facilitate communication between researchers conducting AI evaluations of societal impact and stakeholders beyond the scientific community.}
    \label{fig:risk_report}
\end{figure*}

As noted in \citet{aven2025uncertainty}, there is a strong need for empirical evidence on the best ways to communicate the results of risk assessments which characterize severity, stochastic uncertainty, and epistemic uncertainty. Right now, most common approaches report a combination of consequences, uncertainties, and statements about the strength of knowledge available to the evaluator used to inform their assessments \citep{aven2025uncertainty}. There have been a few proposals of ways to incorporate uncertainty into risk diagrams or other similar ways of visualizing risk characterizations \citep{goerlandt2016assessment}, but they are difficult to interpret and understand for non-domain experts or laypeople. Therefore, inspired by the use of model, dataset, and audit cards to communicate essential information about these technological artifacts, we propose AI Risk Report Cards as a way to communicate risks identified from scholarly or scientific AI evaluations to broad audiences. An example of a possible risk report card based on the re-analyzed case study of \citet{wilson2024gender} is shown in Figure \ref{fig:risk_report}. 

There are six components in our AI Risk Report Card proposal. First, it is necessary to include contextual information about the \textbf{Activity} of interest (including what systems are evaluated, for what tasks, and in what domain) and the associated \textbf{Consequence} (including a description of the possible outcomes evaluated and values implicated). Next, more details about the consequence(s) should be given, including descriptions of the \textbf{Severity} and \textbf{Stochastic Uncertainty}. We suggest describing high-level results in a way that non-experts can understand (e.g., by including scale endpoints as references and minimizing the use of technical language) and summarizing these results using scales describing whether associated quantities are low, medium, high, etc. relative to other possible values. The \textbf{Epistemic Uncertainty} can be described using statements about the strength of knowledge underlying the assessment using criteria described in \citet{flage2009expressing} and reproduced in the appendix. Finally, the overall \textbf{Risk Level Assessment} can be summarized by using the qualitative criteria in \citet{aven2017improving} and reproduced in the appendix as a coarse summary of the information presented elsewhere on the card. While this structure of the AI Risk Report Card was developed in order to portray the essential components of AI risk assessment and evaluation in a succinct and easily understandable manner, its utility is untested. As risk science and AI evaluation practice mature, we hope that further empirical investigation and refinement will be undertaken to validate the card and improve its efficacy. 

\section{Discussion}

In this paper, we presented an argument that scientists and researchers investigating the potential societal impacts of AI should engage more deeply with the field of risk science, which aims to provide justified knowledge about how to conduct, use, and communicate risk analyses \citep{aven2025history}. As the stakeholders of AI societal impact evaluations extend far beyond the scientific community, incorporating the principles of risk analysis can improve scientists' ability to make their work useful and understandable for others affected by AI proliferation. We demonstrated this in two ways: first, by conducting a literature review of fairness evaluations in AI hiring tasks and finding that most overlook critical aspects of risk such as the stochastic and epistemic uncertainty surrounding consequence characterization and measurement. Second, we demonstrated how risk principles could be incorporated into a typical AI evaluation and how this could increase the utility and interpretability of results, especially when summarized in the proposed AI Risk Report Card. 

This work is in line with broader calls to improve the standards for conducting AI evaluations. For example, \citet{stahl2023systematic} argue that AI impact assessments are very disparate and lack shared frameworks for evaluation methods, analysis, and reporting. Risk science aims to identify generic principles of risk analysis regardless of domain, which could potentially address AI evaluation challenges related to generalizability and robustness. Another critique is of AI impact assessments is that they lack real-world grounding and validity \citep{blodgett2021stereotyping}---while risk science cannot directly fix this, it can provide structure to characterizing and qualifying results that arise from evaluations with validity challenges through the use of epistemic uncertainty. 

Future work will be needed to assess our initial proposal to incorporate risk science into AI evaluation more deeply and address its limitations, ideally opening a wide potential research agenda. For example, we situated our literature survey and case study firmly within the domain of fairness evaluation for AI hiring tasks, but it is important to determine how well our findings can generalize to other domains, tasks, and societal impacts. Because one of the primary aims of risk science is to produce general principles, we hypothesize it can apply to AI evaluations more broadly; future work should determine whether any evaluations are incompatible with the principles outlined here. Additionally, we looked primarily at quantitative methods of risk assessment, but additional qualitative assessment strategies can also be valuable, particularly when it comes to characterizing epistemic uncertainties \citep{tiusanen2017qualitative} (and the nature of uncertainty more broadly). Finally, understanding how AI risks are perceived and how risk understanding can be facilitated through high-quality scientific risk analysis and communication is a key area for future work, which will enable greater engagement between researchers and those most directly impacted by their work throughout society.

\section{Conclusion}

This work represents an initial step into a larger research agenda incorporating risk science into scientific assessments of AI models' potential for positive or negative impact by motivating the necessity and benefits of such a convergence using a literature review and case study of fairness evaluation of AI used for hiring or employment tasks. We show that current practices overlook key aspects of risk such as stochastic and epistemic uncertainties, leading to incomplete and potentially misleading results. We offer suggestions for how to use risk science to improve both the practice of AI fairness evaluation and the communication of its results via the AI Risk Report Card, with the intention that understanding risk better will afford better strategies to control and mitigate risk and harm. As the societal impact of AI is something that concerns audiences far beyond the scientific community, it is critical to perform high-quality evaluations and effectively communicate their results so laypeople can incorporate the information into their own practices and decisions surrounding AI.

\section*{Acknowledgments}

We are grateful to the anonymous reviewers for their constructive feedback. This work was supported in part by the Schmidt Sciences Award on AI and Advanced Computing, through the Science of Trustworthy AI program; the U.S. National Science Foundation (NSF) CAREER Award 2337877; and the University of Washington Tech Policy Lab. Any opinions, findings, and conclusions or recommendations expressed in this material are those of the authors and do not necessarily reflect those of supporters. 

\bibliography{aaai2026}

\include{appendix}

\end{document}

%% file: appendix.tex
\appendix
\onecolumn

\section{Qualitative Criteria for Assessing Strength of Knowledge (Reproduced from \citet{flage2009expressing})} \label{sec:SoK_criteria}

\begin{table}[h!]
\begin{tabular}{@{}llll@{}}
\toprule
\multicolumn{1}{c}{Uncertainty} & \multicolumn{1}{c}{\begin{tabular}[c]{@{}l@{}}Strength of \\Knowledge\end{tabular}} & \multicolumn{1}{c}{Criteria} & \multicolumn{1}{c}{Conditions} \\ \midrule
\rowcolor[HTML]{DAE8FC} 
\cellcolor[HTML]{DAE8FC} & \cellcolor[HTML]{DAE8FC} & \cellcolor[HTML]{DAE8FC} & \begin{tabular}[c]{@{}l@{}}The phenomena involved are well understood; the models \\ used are known to give predictions with the required accuracy\end{tabular} \\
\rowcolor[HTML]{DAE8FC} 
\cellcolor[HTML]{DAE8FC} & \cellcolor[HTML]{DAE8FC} & \cellcolor[HTML]{DAE8FC} & Much reliable data are available \\
\rowcolor[HTML]{DAE8FC} 
\cellcolor[HTML]{DAE8FC} & \cellcolor[HTML]{DAE8FC} & \cellcolor[HTML]{DAE8FC} & The assumptions made are seen as very reasonable \\
\rowcolor[HTML]{DAE8FC} 
\multirow{-4}{*}{\cellcolor[HTML]{DAE8FC}Low} & \multirow{-4}{*}{\cellcolor[HTML]{DAE8FC}Strong} & \multirow{-4}{*}{\cellcolor[HTML]{DAE8FC}All conditions are fulfilled} & There is broad agreement among experts \\
\rowcolor[HTML]{D9EAD3} 
\cellcolor[HTML]{D9EAD3} & \cellcolor[HTML]{D9EAD3} & \cellcolor[HTML]{D9EAD3} & \begin{tabular}[c]{@{}l@{}}Example: The phenomena involved are well understood, but \\ the models used are considered simple\end{tabular} \\
\rowcolor[HTML]{D9EAD3} 
\multirow{-2}{*}{\cellcolor[HTML]{D9EAD3}Medium} & \multirow{-2}{*}{\cellcolor[HTML]{D9EAD3}Moderate} & \multirow{-2}{*}{\cellcolor[HTML]{D9EAD3}\begin{tabular}[c]{@{}l@{}}Conditions between low and high \\ are fulfilled to varying degrees\end{tabular}} & Example: Some reliable data are available \\
\rowcolor[HTML]{FFF2CC} 
\cellcolor[HTML]{FFF2CC} & \cellcolor[HTML]{FFF2CC} & \cellcolor[HTML]{FFF2CC} & \begin{tabular}[c]{@{}l@{}}The phenomena involved are not well understood; models\\ are non-existent or known to give poor predictions\end{tabular} \\
\rowcolor[HTML]{FFF2CC} 
\cellcolor[HTML]{FFF2CC} & \cellcolor[HTML]{FFF2CC} & \cellcolor[HTML]{FFF2CC} & Data are not available, or are unreliable \\
\rowcolor[HTML]{FFF2CC} 
\cellcolor[HTML]{FFF2CC} & \cellcolor[HTML]{FFF2CC} & \cellcolor[HTML]{FFF2CC} & The assumptions made represent strong simplifications \\
\rowcolor[HTML]{FFF2CC} 
\multirow{-4}{*}{\cellcolor[HTML]{FFF2CC}High} & \multirow{-4}{*}{\cellcolor[HTML]{FFF2CC}Strong} & \multirow{-4}{*}{\cellcolor[HTML]{FFF2CC}At least one condition is fulfulled} & There is lack of agreement/consensus among experts \\ \bottomrule
\end{tabular}
\end{table}

\section{Qualitative Criteria for Relative Risk Ranking (Reproduced from \citet{aven2017improving})} \label{sec:rank_criteria}

\begin{table}[h!]
\centering
\begin{tabular}{@{}ll@{}}
\toprule
\multicolumn{1}{c}{Rank} & \multicolumn{1}{c}{Criteria} \\ \midrule
\rowcolor[HTML]{C9DAF8} 
Very high risk & \begin{tabular}[c]{@{}l@{}}Potential for extreme consequences, relatively large associated \\ probability of such consequences and/or significant uncertainty\\ (relatively weak background knowledge)\end{tabular} \\
\rowcolor[HTML]{D9EAD3} 
High risk & \begin{tabular}[c]{@{}l@{}}The potential for extreme consequences, relatively small \\ associated probability of such consequences and moderate or\\ weak background knowledge\end{tabular} \\
\rowcolor[HTML]{FFF2CC} 
Moderate risk & \begin{tabular}[c]{@{}l@{}}Between low and high risk. For example, the potential for\\ moderate consequences, and weak background knowledge.\end{tabular} \\
\rowcolor[HTML]{F4CCCC} 
Low risk & No potential for serious consequences \\ \bottomrule
\end{tabular}
\end{table}

\section{Fairness Evaluation Equations and Associated Studies (Reproduced from \citet{fabris2025fairness}} \label{sec:fabris}

\begin{table*}[]
\tiny
\centering
\setlength{\tabcolsep}{4pt} %
\renewcommand{\arraystretch}{1.3} %
    \begin{tabular}{@{}lll@{}}
        \toprule
        \textbf{Measure} & \textbf{Equation} & \textbf{Studies} \\ \midrule
        \rowcolor[HTML]{D9EAD3} 
        Skew@k & $\skewk_g = \log \left ( \frac{N_g^k/k}{D_g} \right ) $ & \citep{geyik2019fairness}
         \\ [.5cm] 
         \rowcolor[HTML]{D9EAD3}
        \begin{tabular}[c]{@{}l@{}}Normalized Discounted\\Cumulative Kullback-Leibler \\ Divergence (NDKL)\end{tabular} & $\ndkl = \frac{1}{Z} \sum_{k=1}^K \frac{1}{\log_2(k+1)} d_{\text{KL}}(D, N^k)$ & \begin{tabular}[c]{@{}l@{}}\citep{geyik2019fairness}\\\citep{arafan2022end}\end{tabular} \\ [.5cm]  
        \rowcolor[HTML]{D9EAD3} 
        Disparate Impact (DI) & \begin{tabular}[c]{@{}l@{}}$\di = \frac{\ming N_g^k / N_g}{\maxg N_{{g}}^k / N_{{g}}}$ \\ $= \frac{\ming \Pr(\hat{y}=1 | s=g)}{\maxg \Pr(\hat{y}=1 | s={g})}$\end{tabular} &
        \begin{tabular}[c]{@{}l@{}}\citep{booth2021bias}\\ \citep{burke2021fair} \\
        \citep{kochling2021highly}\\
        \citep{wilson2021building} \\
        \citep{deshpande2020mitigating}\end{tabular} \\ [.1cm] 
        \rowcolor[HTML]{D9EAD3} 
        Demographic Disparity (DD) & \begin{tabular}[c]{@{}l@{}}$\dd = N_g^k / N_g -  N_{\stcomp{g}}^k / N_{\stcomp{g}}$ \\ $= \Pr(\hat{y}=1 | s=g) - \Pr(\hat{y}=1 | s=\stcomp{g})$\end{tabular} & \citep{rus2022closing}\\ [.1cm] 
        \rowcolor[HTML]{D9EAD3} 
        \begin{tabular}[c]{@{}l@{}}Representation in Positive \\ Predictive Rate (RPPR)\end{tabular} & \begin{tabular}[c]{@{}l@{}}$\rpp_g = \Pr(s=g|\hat{y}=1)$ \\ $\text{RPPD} = \rpp_g - \rpp_{\stcomp{g}} = 2\rpp_g -1$\end{tabular} & \begin{tabular}[c]{@{}l@{}}\citep{ali2019discrimination}\\\citep{imana2021auditing}\\\citep{pena2020bias}\end{tabular}
        \\ [.3cm] 
        \rowcolor[HTML]{D9EAD3} 
        \begin{tabular}[c]{@{}l@{}}True-Positive Rate \\ Difference (TPRD)\end{tabular} & \begin{tabular}[c]{@{}l@{}}$\tpr_g = \Pr(\hat{y} = 1 | y=1, s=g)$ \\ $\tprd = \tpr_g - \tpr_{\stcomp{g}}$\end{tabular} &
        \begin{tabular}[c]{@{}l@{}}\citep{dearteaga2019bias}\\\citep{hemamou2022delivering}\\\end{tabular}\\ [.3cm] 
        \rowcolor[HTML]{D9EAD3} 
        False-Negative Rate Ratio (FNRR) & \begin{tabular}[c]{@{}l@{}}$\fnr_g = \Pr(\hat{y} = 0 | y=1, s=g)$ \\ $\fnrr = \frac{\fnr_g}{\fnr_{\stcomp{g}}}$\end{tabular} & \citep{kochling2021highly}\\ [.1cm] 
        \rowcolor[HTML]{D9EAD3} 
        \begin{tabular}[c]{@{}l@{}}Extended Equality of \\ Opportunity (ExEO)\end{tabular} & \begin{tabular}[c]{@{}l@{}}\begin{small}$\xeo = \max_x [ | \Pr(f_{\text{soft}}(x) \leq t | y=1, s=g) -$\end{small} \\ \begin{small}$\Pr(f_{\text{soft}}(x) \leq t | y=1, s=\stcomp{g})|]$\end{small}\end{tabular} & \citep{nandy2022achieving} \\ [.3cm] 
        \rowcolor[HTML]{D9EAD3} 
        \begin{tabular}[c]{@{}l@{}}Discounted Representation \\ Difference (DRD)\end{tabular} & \begin{small}$\drd = \sum_{k=1}^K \frac{1}{\log_2(k+1)} [\mathds{1}(s_{\tau(k)} = g) - \mathds{1}(s_{\tau(k)} = \stcomp{g})]$\end{small} & \citep{zhang2022are} \\ [.3cm] 
        \rowcolor[HTML]{D9EAD3} 
        Score KL Divergence (SKLD) & $\text{SKL} = \text{KL}(D^f_{g}, D^f_{\stcomp{g}})$ & \citep{pena2020bias} \\ [.1cm] 
        \rowcolor[HTML]{D9EAD3} 
        Log Rank Regression (LRR) & \begin{tabular}[c]{@{}l@{}}$\log (\tau^{-1}(i)) = \beta_x x_i + \beta_s s_i + \mu + \epsilon$ \\ $\lrr = \hat{\beta}_s$\end{tabular} & \citep{chen2018investigating} \\ [.1cm] 
        \rowcolor[HTML]{D9EAD3} 
        \begin{tabular}[c]{@{}l@{}}Root Mean Square of TPRD\\(RMS)\end{tabular} & $\rms = \sqrt{\frac{1}{|\mathcal{S}|} \sum_{g \in \mathcal{S}} \tprd_g^2}$ & \citep{hemamou2022delivering} \\ [.3cm] 
        \rowcolor[HTML]{D9EAD3} 
        Mean Error Difference (MED) & \begin{tabular}[c]{@{}l@{}}\begin{small}$\med = \maxg \frac{1}{N_g} \sum_{i \in g} (y_i-f_{\text{soft}}(x_i))$\end{small} \\ \begin{small}$- \ming \frac{1}{N_g} \sum_{i \in g} (y_i-f_{\text{soft}}(x_i))$\end{small}\end{tabular} & \citep{singhania2020grading}\\ \midrule
        \rowcolor[HTML]{C9DAF8} 
        Mean Absolute Error (MAE) & \begin{small}$\text{MAE} = \frac{1}{N_g} \sum_{i \in g} |y_i-f_{\text{soft}}(x_i)| - \frac{1}{N_{\stcomp{g}}} \sum_{i \in \stcomp{g}} |y_i-f_{\text{soft}}(x_i)|$\end{small} &
        \begin{tabular}[c]{@{}l@{}}\citep{yan2020mitigating}\\\citep{singhania2020grading}\\\end{tabular}
        \\  [.1cm] 
        \rowcolor[HTML]{C9DAF8} 
        \begin{tabular}[c]{@{}l@{}}Balanced Classification \\ Rate Difference (BCRD)\end{tabular} & \begin{tabular}[c]{@{}l@{}}$\text{BCR}_g = \frac{\tpr_g + \tnr_g}{2}$ \\ $\bcrd = \text{BCR}_g - \text{BCR}_{\stcomp{g}}$\end{tabular} & \citep{kochling2021highly}\\ [.3cm]
        \rowcolor[HTML]{C9DAF8} 
        \begin{tabular}[c]{@{}l@{}}Mutual Information Difference \\ (MID)\end{tabular} & \begin{tabular}[c]{@{}l@{}}$\text{MI}_g = \text{MI}_{\{i \in g\}}(\hat{y}_i,y_i)$ \\ $\midif = \text{MI}_g - \text{MI}_{\stcomp{g}}$\end{tabular} & \citep{kochling2021highly}\\
        \rowcolor[HTML]{D9D2E9}  \midrule
        Salary Difference (SaD)& \begin{tabular}[c]{@{}l@{}}\begin{small}$f(x_i) : \text{closest job match}; W(f(x_i)) : \text{avg wage}$\end{small} \\ \begin{small}$\sd = \frac{1}{N_g}\sum_{i \in g} W(f(x_i)) - \frac{1}{N_{\stcomp{g}}}\sum_{i \in \stcomp{g}} W(f(x_i))$\end{small}\end{tabular}  & \citep{rus2022closing} \\ \midrule
        \rowcolor[HTML]{F4CCCC} 
        Gender Bias Score (GBS)& \begin{small}$\gbs = \text{sign}(x_m-x_f) \cdot \max \left \lbrace \frac{x_m-x_f}{x_m}, \frac{x_f-x_m}{x_f} \right \rbrace$\end{small} & \citep{hu2022balancing} \\ \midrule
        \rowcolor[HTML]{FFF2CC} 
        \begin{tabular}[c]{@{}l@{}}Sensitive Area Under the \\ Receiver Operating \\ Characteristic Curve (sAUC)\end{tabular} & $\sauc = \text{AUC}(h_{\text{soft}}(x))$ &
        \begin{tabular}[c]{@{}l@{}}\citep{parasurama2022gendered2}\\
        \citep{parasurama2021degendering}\\
        \citep{booth2021bias}\\
        \citep{hemamou2021judge}\\
        \citep{rus2022closing}\\
        \citep{hemamou2022delivering}\\
        \end{tabular}
        \\ [.5cm]
        \rowcolor[HTML]{FFF2CC} 
        Ground Truth Regression (GTR)& \begin{tabular}[c]{@{}l@{}}$\log \left ( \frac{y_i}{1-y_i} \right ) = \beta_x x_i + \beta_s s_i + \mu + \epsilon$ \\ $\gtr = \beta_s$\end{tabular} & \citep{parasurama2022gendered2}\\ [.15cm] 
        \rowcolor[HTML]{FFF2CC} 
        \begin{tabular}[c]{@{}l@{}}Mutual Information \\ Amplification (MIA)\end{tabular} & $\mia = \text{MI}(\hat{y},s) - \text{MI}(y,s)$ & \citep{yan2020mitigating} \\ \bottomrule
    \end{tabular}
    \caption{Summary of the AI hiring fairness metrics listed in \citet{fabris2025fairness} and the studies which used them. The colors of rows correspond to the different kinds of fairness identified by \citet{fabris2025fairness}: \colorbox{MyGreen}{Outcome}, \colorbox{MyBlue}{Accuracy}, \colorbox{MyPurple}{Impact}, \colorbox{MyRed}{Representational}, and \colorbox{MyYellow}{Process}.}
    \label{tab:fabris}
\end{table*}

\newpage
\section{Supplemental Results}

\begin{figure*}[h!]
    \centering
    {\includesvg[width=0.32\textwidth]{Images/case_DI_distribution_e5.svg}}
    {\includesvg[width=0.32\textwidth]{Images/case_DI_YN_distribution_e5.svg}} 
    {\includesvg[width=0.32\textwidth]{Images/case_DIsig_YN_distribution_e5.svg}}
    \caption{[A] The distribution of observed Disparate Impact (DI) scores when evaluating white men's vs. Black men's resumes using e5. The red line marks the sample average; the two dashed black lines represent different possible DI thresholds at 0.666 and 0.800. [B] The number of resume screening scenarios that are classified as discriminatory vs. non-discriminatory when the DI threshold is 0.800 (based on the 80\% rule). [C] The number of resume screening scenarios which are classified as discriminatory vs. not when the DI threshold is 0.666 (based on statistical significance).}
    \label{fig:case_results_e5}
\end{figure*}

\begin{figure*}[h!]
    \centering
    {\includesvg[width=0.32\textwidth]{Images/case_DI_distribution_SFR.svg}}
    {\includesvg[width=0.32\textwidth]{Images/case_DI_YN_distribution_SFR.svg}} 
    {\includesvg[width=0.32\textwidth]{Images/case_DIsig_YN_distribution_SFR.svg}}
    \caption{[A] The distribution of observed Disparate Impact (DI) scores when evaluating white men's vs. Black men's resumes using SFR. The red line marks the sample average; the two dashed black lines represent different possible DI thresholds at 0.666 and 0.800. [B] The number of resume screening scenarios that are classified as discriminatory vs. non-discriminatory when the DI threshold is 0.800 (based on the 80\% rule). [C] The number of resume screening scenarios which are classified as discriminatory vs. not when the DI threshold is 0.666 (based on statistical significance).}
    \label{fig:case_results_SFR}
\end{figure*}